\documentclass[12pt]{article}
\usepackage[margin=1in]{geometry}
\usepackage{setspace}
\usepackage{graphicx}
\usepackage[natbibapa]{apacite}
\usepackage{hyperref}
\usepackage{booktabs}
\usepackage{tabularx}
\usepackage{array}
\usepackage{authblk}
\usepackage{fancyvrb}
\usepackage{float}
\usepackage{placeins}
\usepackage{xcolor}
\usepackage{caption}
\title{Fact-check? True!: Tracking Verification Requests and the Absence of Truth Verdicts across Correction Systems}
\author[1]{Morgan Wack}
\author[2,3]{Patrick Warren}
\affil[1]{Department of Communication \& Media Research, University of Zurich}
\affil[2]{Media Forensics Hub, Clemson University}
\affil[3]{John E. Walker Department of Economics, Clemson University}
\date{\today}

\begin{document}
\shortcites{altuncu2025factors,aruguete2024confirmation,batailler2022signal,bobek2026grok,ecker2022,hoes2024interventions,lee2023factchecking,pennycook2020implied,renault2026grok,slaughter2025,su2025climateviz,thorne2018fever,wack2026coverage,wack2026olympics,walter2020,wojcik2022}
\singlespacing
\maketitle

\begin{abstract}
\noindent Misperceptions can run in two directions. Individuals can either believe false information or reject true information, with experimental work finding that debunking efforts are effective at correcting both forms of misperception. Yet despite the efficacy of true corrections, current debunking systems focus almost entirely on the correction of falsehoods, with professional fact-checkers publishing few true verdicts and crowdsourced systems like X's Community Notes program unable, by rule, to display notes verifying true claims. As a result, we do not have a clear understanding of the extent to which overall misperceptions are driven by doubting true information. To answer this question we first draw on a novel dataset of 47,899 queries to X's AI assistant which track users asking for verification on post content. Surprisingly, among questioned posts that received a true or false verdict, Grok judged nearly identical shares as either accurate or inaccurate. Of the questioned posts, more than 83\% were not flagged by Community Notes and only 3.3\% carried a displayed note. Among questioned posts judged accurate by Grok, nearly 90\% received no note at all while fewer than 2\% displayed a note flagging them as misleading. This discrepancy does not seem to have stemmed from a lack of effort, with contributors writing 2,981 notes verifying questioned posts despite there being no way for Community Notes to display them publicly. Alongside further exploration of these findings we discuss both the system- and code-level changes that would be required to effectively address this asymmetric correction paradigm.
\end{abstract}
\clearpage
\doublespacing

\section{Introduction}

\begin{figure}[h]
\centering
\includegraphics[width=0.75\linewidth]{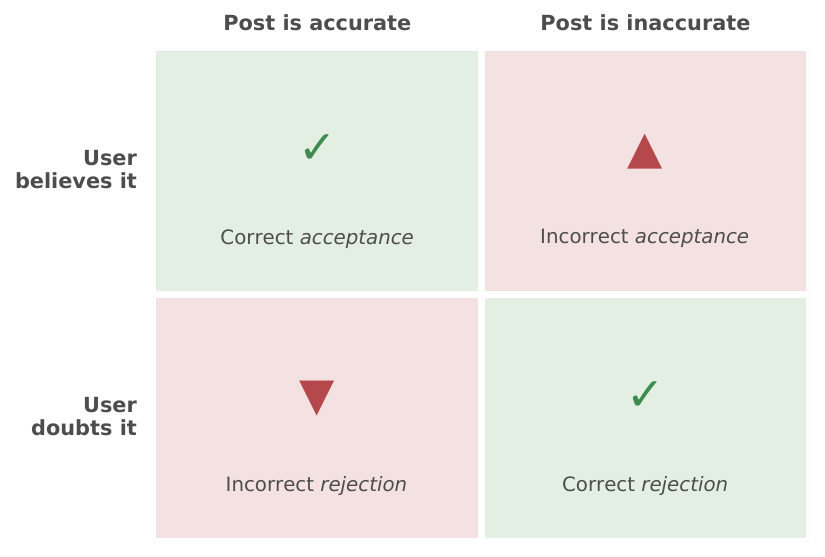}
\caption{The four types of accuracy perception. Columns denote the underlying truth, while rows refer to levels of user belief. Notes flagging a post as misleading address incorrect acceptance (top right). Notes verifying a post as not misleading address incorrect rejection (bottom left), the doubting of an accurate post.}
\label{fig:conceptual}
\end{figure}

Fact-checking has become the most prominent intervention for combating online misinformation. Meta-analyses and multi-country experiments showed that corrections can reduce belief in false claims \citep{walter2020, porter2021global, ecker2022}. Reviews also found that warning labels can reduce both belief in and sharing of false claims \citep{martel2023}. Yet this evidence, and the infrastructure it supports, often focuses on the correction of false information over the promotion of true information that is, nevertheless, questioned by the public. This asymmetric emphasis is misguided. A meta-analysis of 67 experimental articles found that people were worse at recognizing true news than false news \citep{pfander2025spotting}. Figure~\ref{fig:conceptual} presents these two forms of misperception, incorrect acceptance, believing a post that is inaccurate, and incorrect rejection, doubting a post that is accurate. The terms follow signal detection approaches to misinformation, which separate the ability to distinguish true from false information from a general tendency to accept or reject claims \citep{batailler2022signal}. Although fact-checks that verify true claim are rare, studies on their use have found that confirming accurate content can strengthen belief in accurate information \citep{wack2026olympics}.

The absence of these confirmations can be consequential.  Labeling some false headlines increased the perceived accuracy of unlabeled ones \citep{pennycook2020implied}, while in survey experiments, fact-checking, media literacy, and misinformation-coverage interventions reduced belief in false information but also reduced the perceived credibility of factual information \citep{hoes2024interventions}. In a four-country survey experiment, respondents were also more willing to engage with corrections framed as confirmations than as refutations \citep{aruguete2024confirmation}. 

Despite these concerns, evidence on the use of true fact-checks has largely been limited to survey experiments in part due to their comparatively limited real-world application. As an example, one of the most prominent meta-analysis on the efficacy of fact-checks had planned to include favorable fact-checks alongside corrections but set them aside because too few studies existed to analyze \citep[n. 2]{walter2020}. Professional fact-checkers rarely publish true verdicts, with True or Mostly True verdicts accounting for only 11\% of PolitiFact's and 29\% of Snopes' articles \citep{lee2023factchecking}. Crowdsourced systems, which have been proposed as alternatives to fact-checking at scale, perform even worse. X's Community Notes, a crowdsourced fact-checking system \citep{wojcik2022}, allows contributors to write ``not misleading'' verdicts but presents no avenue for these notes to be displayed to users. In short, although incorrect rejections pose real-world consequences that remain addressable in principle, many of the systems designed to correct misperceptions ignore them entirely.

Existing studies of true fact-checks have also been constrained by their reliance on survey experiments, which often use true claims as filler content. As a result, their use cannot tell us how often the content users question is in fact accurate nor how active moderation systems respond to these questions in practice. If incorrect rejections make up only a small share of the doubt users express, the asymmetry described above represents only a minor inefficiency. If, however, they make up a large share of all doubted content, they represent a structural gap in how platforms address misperceptions. To fill these gaps, we ask: What proportion of the content that is doubted online is true and how do active interventions currently address incorrect rejections of accurate content?

To answer these questions we draw on the queries X/Twitter users post expressing uncertainty about social media content. Users routinely reply to posts by asking the platform's AI assistant, Grok, whether the information they contain is true, with verification a common use of Grok on the platform \citep{renault2026grok, bobek2026grok}. Importantly, each of these queries leaves a public record of what was questioned as well as whether an answer provided by Grok. We collected 47,899 of these queries posted between January and February 2026 and joined these posts with publicly released Community Notes data. The addition of the Community Notes data allow us to assess whether contributors wrote notes on these flagged post as well as whether notes were displayed to users. With this data, we are able to estimate the share of questioned posts that were judged to be accurate while observing how Community Notes, the most well-known crowdsourced moderation system, responded to these same claims. 

The paper proceeds as follows: Section~\ref{sec:background} describes the two systems, Sections~\ref{sec:data} and~\ref{sec:methods} describe the corpus and the coding, Section~\ref{sec:results} reports the findings, and Section~\ref{sec:discussion} overviews what the corpus establishes, what it does not, and what the development of a truly symmetric system would require.

\section{Background}
\label{sec:background}

\subsection{Verification Queries to AI Assistants}

Grok, X/Twitter's interactive AI bot, responds to most user queries, and users often ask it to assess the veracity of a post's claims.\footnote{Currently, at times of high demand, responses are limited to X Premium subscribers, but during the period when we collected our data, there was no such restriction.} Users who doubt a post often directly ask ``@grok is this true?'', which prompts the bot to post an answer beneath the query, typically in a matter of seconds. This occurs at scale, with \citet{renault2026grok} counting 1,555,275 English-language fact-checking requests to Grok and Perplexity between February and September 2025. Although these requests typically reflect genuine user doubt, spot-checks suggest not all doubt is sincere, with some users instead tagging the bot to mock a poster. To address this subset, we provide detail on how we deal with these insincere queries in Section~\ref{sec:methods}.

Two recent studies have examined these requests at scale. \citet{renault2026grok} asked who made these requests and tracked the accuracy of the bots' answers, finding that Grok's requesters skewed Republican while Perplexity's skewed Democrat. They also found that, on 188 posts rated by three hired professional fact-checkers, Grok-3's verdicts agreed with a fact-checker 57.9\% of the time, against 59.3\% agreement among the fact-checkers themselves. Relatedly, \citet{bobek2026grok} analyzed 169,137 prompt--reply pairs involving Grok, finding that users call on the bot to reply to specific posts, that the bot's answers often only reach small audiences, and that the posts users query overlap little with those that receive Community Notes. \citet{renault2026grok} also examined the veracity of queried posts and compared LLM judgments with professional fact-checkers and with displayed Community Notes. We extend this work by examining the composition of doubted content, the notes contributors wrote verifying queried posts, and the rule that keeps these notes from being displayed beneath posts.

\subsection{Community Notes and the Display Exclusion}
\label{sec:architecture}

Any X/Twitter user who meets the Community Notes enrollment threshold can write a note, with each note classifying targeted posts as either ``potentially misleading'' or ``not misleading'' alongside open-text explanations and source links. Below, we describe a note classified as ``potentially misleading'' as flagging the post and a note classified as ``not misleading'' as verifying it. Once labeled, other contributors then rate each note as either ``Helpful'', ``Somewhat Helpful'', or ``Not Helpful'' while a matrix-factorization scorer aggregates these ratings \citep{wojcik2022}. Displayed notes appear beneath the targeted post, while notes classified as ``Not Helpful'' are not displayed beneath it.\footnote{X/Twitter's documentation for Community Notes states that ``only notes that indicate a post is `potentially misleading' and earn the status of Helpful are eligible to be displayed on posts'' and that a note claiming a post is not misleading ``will be shown to other raters as additional information, but won't show on the post'' \citep{cn_docs}. The same documentation states that X/Twitter ``plan[s] to enable Helpful statuses for notes marking posts as `not misleading' as we continue to evaluate ranking quality and utility to users.''} These notes were not always excluded, with a status assigned to notes verifying posts until March 2022, when X/Twitter paused them after observing that such notes ``were often rated as `Unhelpful - Tweet doesn't need a note'\,'' \citep{cn_docs}. X/Twitter revised the rating form on October 3, 2022, resumed assigning statuses to these notes in select circumstances, and resumed them more broadly on November 25, 2022. Helpful status has remained unavailable to them \citep{cn_docs}. 

\section{Data}
\label{sec:data}

\subsection{Verification Threads}

To assemble our dataset, we collected a 10\% random sample of queries on X/Twitter addressed to Grok that included our informed keyword corpus between 1 January and 28 February 2026.\footnote{The population from which we collected included all messages satisfying the following keyword strings ``@grok'' AND NOT (author:grok) AND (messageType: ``X Reply'') AND (``factcheck this'' OR ``fact-check this'' OR ``is it true'' OR ``is this true'' OR ``is this correct'')''.} These replies come from a Sprinklr export, which records the query's tweet ID, author, timestamp, and the tweet ID of the post to which the query replied. We then gathered full reply threads, including the original claim and Grok's response with their own timestamps, tweet ids, and message text. We recovered Grok replies for 36,053 queries (75.3\% of the export) and the text of 34,691 queried posts (88.5\%).\footnote{The remaining queries were marked during collection as having no reply (4,657), as never attempted (3,655), or as ambiguous (3,533), with one record lacking a status. A missing reply therefore does not establish that Grok did not answer. For 61 recovered replies, the reply's identifier showed that it was created before the query or that its recorded time was off by more than five seconds. We treated these answers as unavailable and kept their queries in the sample.} The remaining 4,495 posts (11.5\%) had either been deleted, made private, or removed by the time of our data collection in August 2026. A second fetch of the original posts on which the Grok queries was made recorded whether each was a reply within a conversation or a top-level post.  Of the live posts, only 5,384 (15.5\%) were replies.

To make the comparison between Grok queries and Community Notes tractable, we restricted our analysis to verification queries on top-level posts as Community Notes rarely address replies.\footnote{Across the 5,384 replies in our corpus only one had a displayed note at collection. This restriction had a negligible effect on the results, moving the share of posts Grok judged true from 51.0\% to 51.4\%. Appendix Table~\ref{tab:samples} reports the main results for all live posts, for top-level posts, and for the sincere-only subset.} We also exclude deleted posts. Section~\ref{sec:supply} reports the unrestricted figures alongside the restricted ones.

\subsection{Community Notes}

To supplement our Sprinklr collection, we downloaded the Community Notes public data release from October 1st, 2026, which covers every note written since January 28th, 2021. In total, this release contains 3,133,544 Community Notes, 235,338,530 ratings, and each note's status history.\footnote{As the release postdates the request window by seven months, every queried post had at least this much time to attract a note. The replication repository lists each release file with its size and checksum, together with the scripts that process it.} Finally, we joined the queried posts to these notes to create our set of community and crowd-sourced queries and corrections.

\subsection{Final Sample}

Table~\ref{tab:corpus} lists each restriction implied between the initial export and the creation of our final sample. We drop replies to make the comparison more even as Community Notes rarely addresses replies and drop deleted posts to remove unrecoverable content. No queries are excluded for their wording or intent. The final sample contains 36,675 queries addressed to 29,294 top-level posts. We recovered Grok's answer for 30,390 of these queries, which gave a directional verdict (true/false) on 23,572 of them.

\begin{table}[!htbp]
\centering
\small
\begin{tabular}{lrrrr}
\toprule
& \textbf{Queries} & \textbf{\%} & \textbf{Posts} & \textbf{\%} \\
\midrule
\multicolumn{5}{l}{\textit{Sample construction}} \\
Sprinklr export & 47,899 & 100.0 & 39,186 & 100.0 \\
Queried post still online & 42,107 & 87.9 & 34,691 & 88.5 \\
Queried post is top-level (final sample) & 36,675 & 76.6 & 29,294 & 74.8 \\
\midrule
\multicolumn{5}{l}{\textit{Subsets of the final sample used in Section~\ref{sec:demand}}} \\
Queries with a recovered Grok answer & 30,390 & 82.9 & & \\
Queries with a true or false verdict (on 19,042 posts) & 23,572 & 64.3 & & \\
\bottomrule
\end{tabular}
\caption{Details on the construction of our final sample. Percentages in the upper section are shares of the Sprinklr export. The lower section presents shares of the final sample's queries.}
\label{tab:corpus}
\end{table}

\section{Methods}
\label{sec:methods}

\subsection{Primary Coding}

The LLM Claude (claude-haiku-4-5) was used to coded Grok's answers to user queries, where one existed. This coding process was structured to identify Grok's verdict (either true, false, mixed, or declined), dispute type (the truth of a stated claim, the attribution of a quote, the authenticity of an image or video), and a sourcing category (a link or named source, an unnamed gesture toward reports or searches, or no warrant).\footnote{Appendix~\ref{app:prompts} reproduces the coding instructions.} Two humans coded Grok's reply on 100 randomly drawn answered queries into the same four categories. Agreement was high when both coders assigned a true or false verdict, but lower across the full set of categories.\footnote{Across all four categories, agreement was 84\% between the two human coders ($\kappa = 0.76$) and 74\% between each human coder and Claude ($\kappa = 0.60$--$0.61$), based on all 100 replies. When both members of each pair coded a reply true or false, the human coders agreed on 99\% ($n = 72$, $\kappa = 0.97$), and each agreed with Claude on 96\% to 97\% ($n = 66$ and $68$, respectively; $\kappa = 0.91$--$0.94$). Nearly all disagreements were concentrated in determinations of whether a response was true/false or mixed, which Claude assigned more often than either human coder. As a result, each coder's true share was higher on average, suggesting that the true share reported below is a conservative estimate (Appendix~\ref{app:humancheck}).} \footnote{A second coding pass asked whether each query was a sincere request to verify the post. Claude coded 24.4\% of all queries as rhetorical or as containing no question, a far larger share than the roughly 5\% of answers that \citet{renault2026grok} excluded as humorous or the 9.9\% of all prompts that \citet{bobek2026grok} coded as humor, although these shares measure different units. To ensure this isn't responsible for our results we report every result on the full set of queries in the main text and use this coding only as a sensitivity check. Restricting to the 31,696 queries it coded sincere moves the share of posts Grok judged true from 51.4\% to 53.2\% and changes no other result by more than 2\% (Appendix Table~\ref{tab:samples}).}

\subsection{Secondary Model Agreement}

Throughout the paper, accurate and inaccurate notations refer to Grok's verdicts, which reflect the answers received by the users who queried Grok on the platform. Each post was assigned Grok's first true or false verdict, ordered by query time. Posts with no true or false verdict were assigned the first available mixed or declined verdict. Other reasonable definitions gave a true share within one point of this primary outcome (Appendix Table~\ref{tab:definitions}). As a check on whether these verdicts were not the product of a model that either affirms or denies all queries, we drew a stratified sample of 600 query records, 200 from each of Grok's true, false, and mixed verdicts, from an earlier subset of queries coded as sincere requests. The records concerned 594 distinct posts, 64 of them replies. Perplexity's search-grounded model (sonar) then judged whether the central claim of each post was accurate. Among the records it could resolve, Perplexity agreed with 84.2\% of Grok's true verdicts ($n = 177$) and 79.8\% of its false verdicts ($n = 178$). Appendix~\ref{app:adjudication} reports the cross-tabulation and Appendix Table~\ref{tab:definitions} reports the true share that results from treating Perplexity's judgments as a reference.

\subsection{Topic Coding}

To see whether doubt, verification, and/or the crowd's response vary by subject, we assign each questioned post a broad topic with a transparent keyword classifier written for the requested corpus which scored the post text and the request against term lists for ten topics before assigning the best-scoring one (or `Other' where none matched). Appendix~\ref{app:topics} lists the topics and their terms. 

\subsection{Public Expectations}

To assess public expectations about Grok's verdicts, we included a question in Verasight's APSA omnibus survey, which was assigned to 1,000 U.S. adults and fielded between the 25th and 30th of September 2026. After introducing Grok as a chatbot used to fact-check social media claims, we asked: ``On average, for all the fact-check questions Grok is asked to judge, for what fraction do you think Grok responds that the claim was `true'?'' Respondents were prompted to select only one of seven ranges of percentages: 0--14\%, 15--29\%, 30--44\%, 45--54\%, 55--69\%, 70--84\%, or 85--100\%. We rely on Verasight's supplied population weights for our analyses.

\section{Results}
\label{sec:results}

Our results describe what each system (Grok or Community Notes) provided to user queries. For Grok, these verdicts were presented as direct reply, with the post's accuracy/inaccuracy coded in relation to the initial post. For Community Notes, the outcomes refer to whether or not contributors wrote an accompanying note on the post as well as whether a note had a ``Helpful'' status at collection and was eligible for public display. 

\subsection{Rates of Doubted Content Accuracy}
\label{sec:demand}

\begin{figure}[!htbp]
\centering
\includegraphics[width=\linewidth]{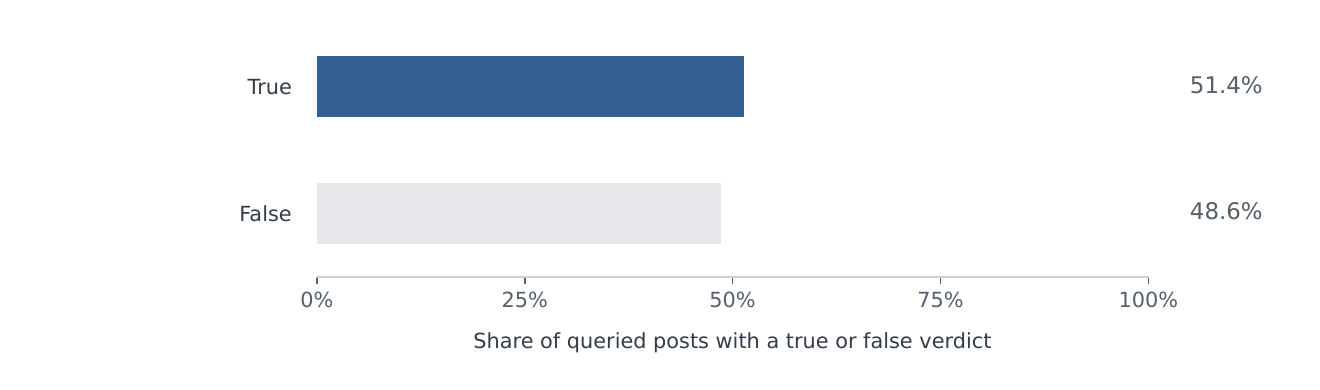}
\caption{Grok's true and false verdicts on the 19,042 queried posts on which it returned one.}
\label{fig:demand}
\end{figure}

Grok judged 51.4\% of the queried posts on which it gave a true or false verdict to be true ($n = 19{,}042$, Figure~\ref{fig:demand}). \footnote{Across all 24,439 posts with an answer, Grok's verdict was true for 40.0\%, false for 37.9\%, mixed for 20.0\%, and declined for 2.1\%. Counting each query separately gave a true share of 46.9\%, and treating Perplexity's judgments on the 600 sampled records as the reference gave 50.2\% (Appendix Table~\ref{tab:definitions}). Among Grok's mixed verdicts, Perplexity found 100 of the 167 claims it could resolve accurate.} In short, readers who doubted a post enough to seek verification were about as likely to have doubted something accurate as something inaccurate, suggesting that roughly half of the demand for verification by X/Twitter users is focused on the types of claims that current systems are ill-equipped to counter.

Survey respondents also expected Grok to judge a substantial portion of its queried claims to be true. Among the survey's weighted sample of U.S. adults, the median estimate of how often Grok judged queried claims to be true was in the percentage band ranging from 45\% to 54\%. This result mirrors our estimates of Grok's assessments, with both these observed verdicts and respondents’ expectations contradicting these correction systems’ focus on false claims.

Two further features of the corpus described what users questioned, which largely came from propositional content. Disputes over a post's facticity accounted for 89.8\% of answered queries, against just 4.6\% aimed at verifying the attribution of a quotation and 5\% regarding whether or not an image or video was authentic. \footnote{Of Grok's answers, only 30.4\% named a source or gave a link, while 59.8\% gestured at unnamed reports or searches and 9.8\% asserted a verdict without accompanying evidence.}

\subsection{Community Notes vs. Grok}
\label{sec:supply}

\begin{figure}[!htbp]
\centering
\includegraphics[width=\linewidth]{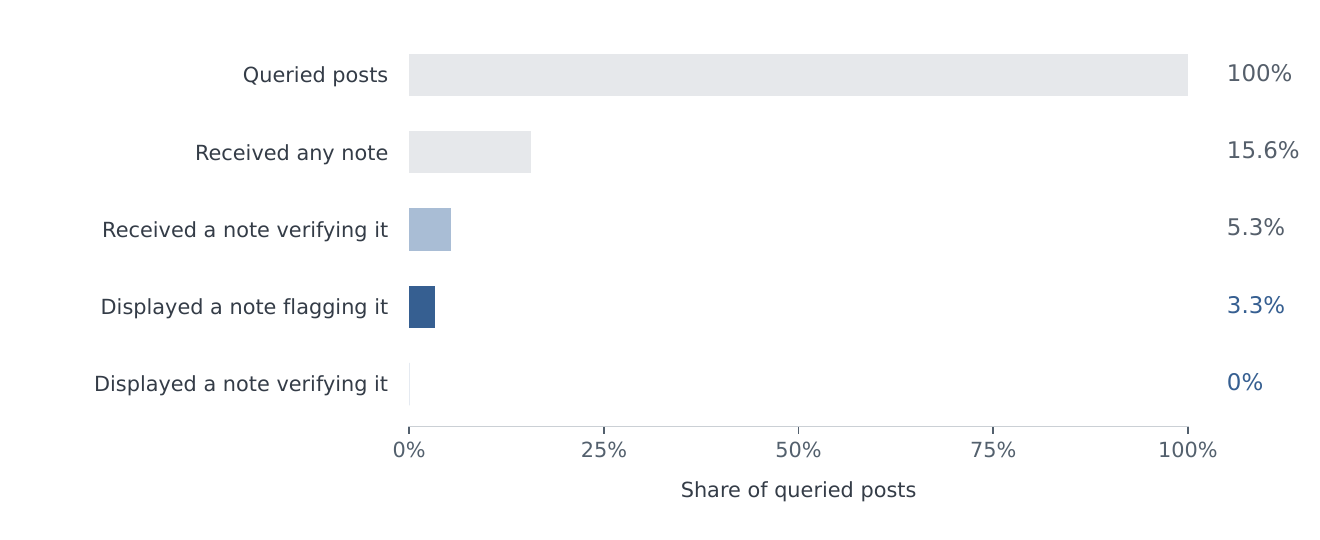}
\caption{Community Notes on the 29,294 queried posts. Percentages use all queried top-level posts as the denominator. Categories overlap, since a post may receive both a note flagging it and a note verifying it. Display refers to status at collection.}
\label{fig:supply}
\end{figure}

Every post in the final sample was questioned by at least one user, enabling us to compare what Community Notes contributors and Grok separately provided to the wider X/Twitter userbase related to the same post content. Of the 29,294 posts queried by Grok users, 24,729 (84.4\%) received no Community Note of any kind. Contributors wrote at least one note on 4,565 (15.6\%) of these posts while a note was only displayed for 968 (3.3\%) of these posts. Weighted by reach, measured as each post's like count, the posts that received any note held 33.6\% of all likes on queried posts and those with a displayed note held 5.3\%, while the share of posts Grok judged to be true held at 50.7\% (Appendix Table~\ref{tab:reach}). On the same posts, Grok answered 30,390 queries, which equates to about thirty Grok answers for every post with a displayed note (Figure~\ref{fig:supply}).\footnote{Grok responses were posted with median delay of 59 seconds (interquartile range 45 to 75).} We also examined whether notes preceded the Grok queries. Among the 4,565 queried posts with a note, the first note preceded the first query 45.5\% of the time, with the median note written half an hour after the Grok query. This suggests that CN contributors and regular users were reacting to the same posts during the same initial window, reflecting evidence on the virality of recent content from the wider fact-checking literature. Among the 968 of these posts that displayed a note at the time of our collection, only 55 (5.7\%) displayed one before the first Grok query. The median displayed note appeared 7.3 hours after the first Grok query and 82.7\% appeared within a day (Appendix Table~\ref{tab:timing}).

Contributors also wrote 2,981 notes verifying 1,557 of the queried posts, 27.3\% of all notes written on them. None was displayed, since the scoring code excludes notes with this classification from display (Section~\ref{sec:architecture}).

\subsection{Outcomes by Accuracy}
\label{sec:cells}

\begin{figure}[!htbp]
\centering
\includegraphics[width=\linewidth]{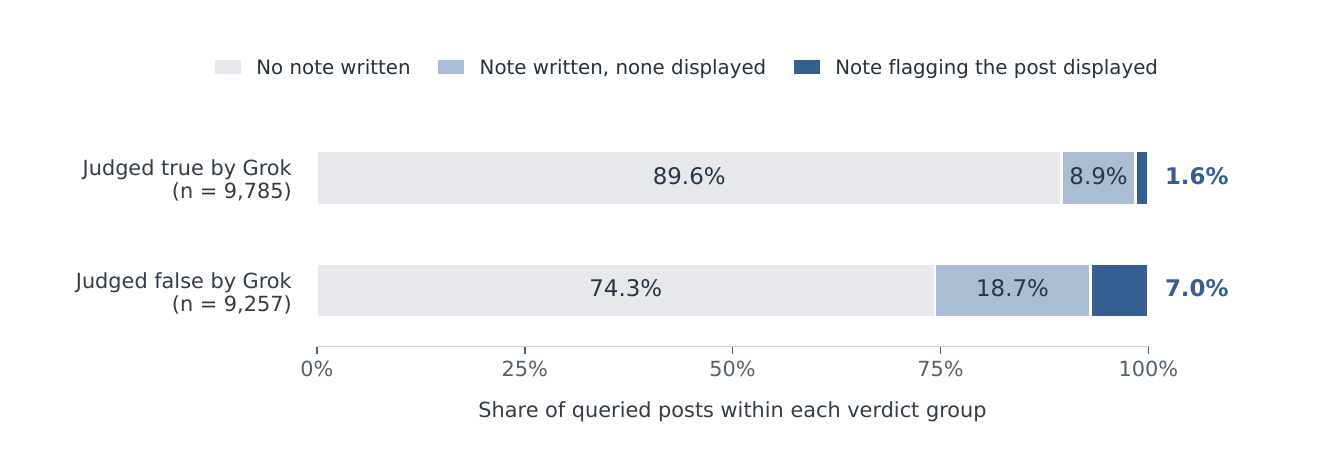}
\caption{Community Notes coverage of queried posts, by Grok's post-level verdict as defined in Section~\ref{sec:methods}. Bars show the shares of top-level posts receiving no note, receiving a note but none displayed, or showing a displayed note flagging them at collection.}
\label{fig:platform}
\end{figure}

Figure~\ref{fig:platform} divides the queried posts by Grok's verdict and by whether a Community Note was displayed ($n = 19{,}042$). Posts judged to be inaccurate received a displayed note 4x as often as posts judged to be accurate (7.0\% against 1.6\%), illustrating that contributors and raters were more likely to note posts when they were inaccurate. Differences were present at both stages: posts Grok judged true were less likely to receive a note flagging them, and those notes were less likely to be displayed. They attracted more ratings but a smaller share of ``Helpful'' ratings, suggesting that raters were more divided.\footnote{Notes flagging the post were written on 10.0\% of posts Grok judged true and 25.4\% of posts judged false; among those receiving such a note, 15.5\% and 27.5\%, respectively, had one displayed. Notes flagging posts judged true attracted a median of 79 ratings, against 45 for notes flagging posts judged false, with 47.8\% and 54.7\% of ratings marked ``Helpful'', respectively. For the full results of these comparisons see Appendix Table~\ref{tab:ratings_verdict}.} 

Few posts in either group, however, received a displayed note, consistent with the limited coverage that \citet{wack2026coverage} documented for professional fact-checking. Among posts Grok judged to be true, 89.6\% received no note of any kind.\footnote{Of the displayed notes flagging posts Grok judged to be true, 37.4\% were coded as addressing framing, context, outdated information, or subsidiary details without disputing the central claim (Appendix~\ref{app:objection}).} 

\subsection{True Verdicts across Verification Systems}
\label{sec:systems}

\begin{figure}[h]
\centering
\includegraphics[width=\linewidth]{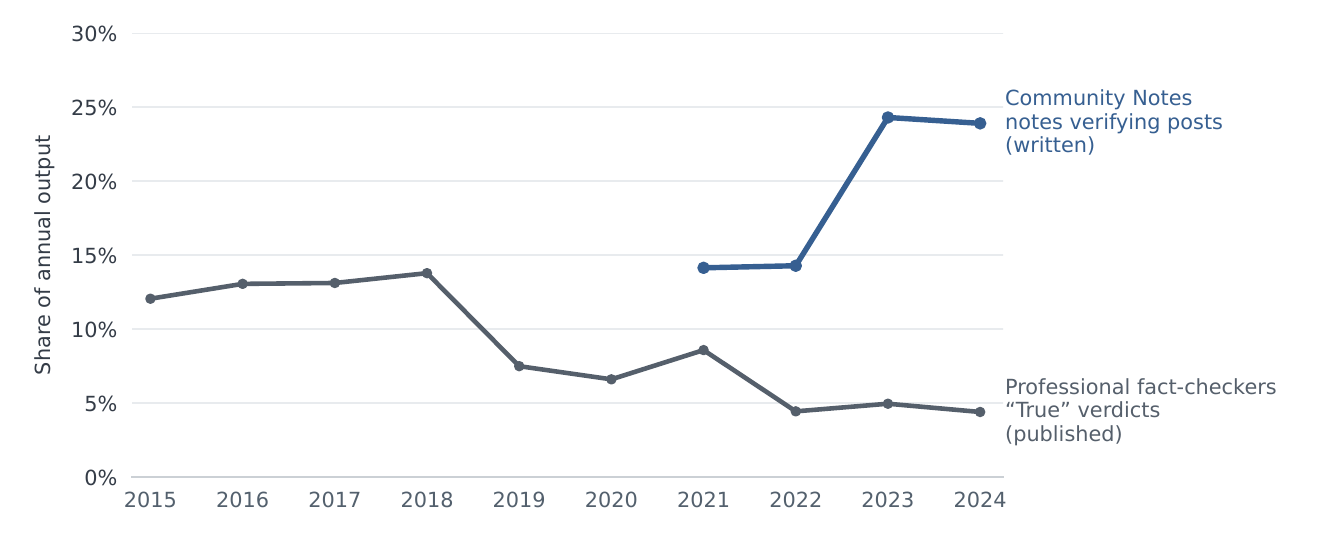}
\caption{Annual shares of true verdicts published by professional fact-checkers and not-misleading notes written by Community Notes contributors. Professional shares use FACTors checks classified as true, partially true, false, or misleading; Community Notes shares use all written notes in the October 1st, 2026 release. Shares pool items within each year, and the organizations represented in FACTors vary over time. The series describe different selected claims and classifications; written notes verifying posts do not measure public display.}
\label{fig:share}
\end{figure}

While roughly half of the posts users questioned were accurate, every verification system that relied on external selection to determine which claims to check favored inaccurate claims, unlike Grok, which responded directly to users' queries (Table~\ref{tab:systems}). Figure~\ref{fig:share} compares the annual composition of professional fact-checkers' and Community Notes contributors' output. The Table reports, for each system, the share of each system's verdicts that determined the selected claims to be true, describing the composition of each system's output.

Grok judged 51.4\% of the queried posts true among posts receiving a true or false answer. By contrast, Community Notes contributors chose which posts to respond to, resulting in their choices favoring the selection of inaccurate content. Notes verifying posts made up 27.3\% of the Community Notes written on the queried posts and 19.6\% of the 3,133,544 notes in the full release. This share ranged from 11.8\% to 28.1\% of monthly outputs since late 2022 (Appendix Figure~\ref{fig:share_monthly}).  \footnote{Two notes verifying posts in the release retained ``Helpful'' status (both from the 2021 pilot), against 267,538 notes flagging posts with that status.} The exclusion applied irrespective of the ratings these notes received. On the queried posts, raters marked notes verifying a post ``Helpful'' more often than notes flagging it as misleading (60.6\% against 50.0\%) while reaching slightly fewer of them, with 82.1\% against 91.6\% receiving at least five ratings (Appendix Table~\ref{tab:ratings}). \footnote{Across the full release the ``Helpful'' share of notes verifying posts remained a few points below that of notes flagging them on average since 2023. X/Twitter's stated reason for pausing their statuses in 2022 was that raters often tagged them as not needed. In the 2022 ratings that tag appeared on 11.2\% of the ``Not Helpful'' ratings given to notes verifying posts and on 49.4\% of those given to notes flagging posts. However, since the revised rating form of October 2022 the tag has not been available for notes verifying posts at all (Appendix Table~\ref{tab:ratings_year}).}

\begin{table}[h]
\centering
\small
\begin{tabular}{lr}
\toprule
\textbf{System} & \textbf{True share} \\
\midrule
Grok's verdicts on the queried posts & 51.4\% \\
Community Notes written on the queried posts & 27.3\% \\
Professional fact-checkers & 8.0\% \\
\bottomrule
\end{tabular}
\caption{Share of each verification system's verdicts on true claims, or not misleading for Community Notes. Grok's denominator includes true/false answers; Community Notes includes all written notes. The professional fact-checker figure comes from FACTors \citep{altuncu2025factors}, 118,112 claims from 117,993 English-language fact-check reports published from 1995 to 2025 by 39 organizations that are signatories of the International Fact-Checking Network or the European Fact-Checking Standards Network. The shares are our calculations from these data.}
\label{tab:systems}
\end{table}

Professional fact-checkers published true verdicts even less often, with 11\% of PolitiFact's and 29\% of Snopes' articles rating their targeted claims either True or Mostly True \citep{lee2023factchecking}. Across the 39 organizations in FACTors \citep{altuncu2025factors}, 8\% of the 115,952 claims with a true, partially true, false, or misleading verdict were rated true, or 11\% when verdicts of Mostly True are included.\footnote{FACTors assigned each organization's original verdict to one of six categories. We used its true category and added the claims whose original verdict was Mostly True, which FACTors placed in its partially true category alongside verdicts such as Half True and Mostly False.} Among the 26 organizations that published at least 500 checks, the median true share was just 1.1\%. True verdicts also became rarer over time, falling from 13\% of output in 2015 to 2018 to 5\% in 2022 to 2024 (Figure~\ref{fig:share}). Among organizations already active by 2018, the share fell to 7\%. \footnote{Recounting with only the original verdicts that the dataset maps to a single normalized rating, 91.5\% of the checks, gave 7.6\%, suggesting that ambiguous verdict mappings do not account for the low true share.} The one outlier in the dataset, the Wisconsin Watch, differs in that it answers assigned reader questions with a yes or a no directly, with 73.7\% of its included checks classified as true in FACTors, showing that true verdicts can constitute a substantial share of a professional fact-checker's output.

Research on professional fact-checking identifies checkability, available evidence and resources, and virality as prerequisites for selecting claims, alongside considerations such as prominence and relevance \citep{cazzamatta2026truth}. These selection criteria provide context for the composition of published verdicts, although our comparisons do not establish which criteria account for the low true share. Grok does not face the same selection choice, since it answers the queries users send it. Its true and false labels also resemble those of automated verification benchmarks, which label claims as supported, refuted, or unverifiable and construct refuted claims by altering true ones \citep{thorne2018fever, su2025climateviz}.

\subsection{Accuracy and Community Notes Coverage by Topic}
\label{sec:topics}

Table~\ref{tab:topics} divides the queried posts by topic. The share of posts judged accurate varied little across the ten named topics, from 49.4\% in religion and identity to 58.3\% in science and climate, while the share receiving a note was more varied. Queried posts about politics and government, crime and law, and health and medicine received a note roughly a fifth of the time, while posts about the economy and sports received a note in only about one in nine cases. The share receiving a displayed note ranged from 1.9\% in sports to 5.1\% in technology. Within every topic, inaccurate posts received displayed notes more often than accurate ones (Appendix Table~\ref{tab:topics_full}). Coverage therefore varied across topics, while inaccurate posts received displayed notes more often across every topic.

\begin{table}[!htbp]
\centering
\small
\begin{tabular}{lrrrr}
\toprule
\textbf{Topic} & \textbf{Posts} & \textbf{Accurate} & \textbf{Any note} & \textbf{Displayed} \\
\midrule
War and geopolitics & 4,530 & 50.8\% & 15.5\% & 3.0\% \\
Politics and government & 4,378 & 52.0\% & 20.2\% & 2.7\% \\
Economy and business & 2,327 & 57.7\% & 11.1\% & 2.8\% \\
Crime, law, and safety & 1,608 & 53.2\% & 20.6\% & 3.7\% \\
Sports & 1,116 & 55.0\% & 11.3\% & 1.9\% \\
Technology and AI & 1,075 & 55.5\% & 15.2\% & 5.1\% \\
Health and medicine & 723 & 49.6\% & 21.7\% & 4.7\% \\
Religion and identity & 668 & 49.4\% & 17.4\% & 3.7\% \\
Entertainment and culture & 441 & 50.2\% & 15.2\% & 4.1\% \\
Science, climate, and environment & 322 & 58.3\% & 18.3\% & 4.3\% \\
Other or unclear & 12,106 & 49.3\% & 14.0\% & 3.5\% \\
\midrule
All & 29,294 & 51.4\% & 15.6\% & 3.3\% \\
\bottomrule
\end{tabular}
\caption{Queried posts by topic. Accuracy corresponds to Grok's post-level verdict under the rule in Section~\ref{sec:methods}, which is the definition used throughout. Any note and Displayed are the shares of posts on which contributors wrote a note and on which a note was displayed. Topics were assigned by a keyword classifier applied to the post and query.}
\label{tab:topics}
\end{table}

\section{Discussion}
\label{sec:discussion}

Efforts to address misperceptions have largely focused on correcting belief in false information, despite evidence that individuals also struggle to recognize information that is accurate \citep{pfander2025spotting}. Through our examination of the content users themselves publicly express doubt, we find that the emphasis of prior attention on false claims does not reflect the patterns of doubted content online, with about half of the posts queried by users judged to be accurate. Given that this pattern held across topics and sample definitions, we read this as evidence that the claims that social media users doubt are, quite broadly, as likely to be accurate as inaccurate.

To illustrate existing biases in the attention paid to these symmetric forms of misperception, our within-platform comparison with Community Notes data showed how this demand is largely unmet. Not only was the display of notes on these queried posts low, current restrictions also ensured that no note verifying a post was ever displayed to the wider public. While contributors wrote notes verifying many of these posts and raters judged them to be helpful, current emphasis on false claims means that even helpful notes on true content are kept from being publicly displayed. 

This same asymmetry in attention appeared beyond Community Notes. While comparison between Grok and Community Notes allowed us to observe what the same audience saw on the same posts, professional fact-checking remains the most common form of debunking used at scale. Within this group, we show that professional fact-checkers also rarely publish true verdicts and have done less often over time. Thus, both volunteer contributors and professional organizations, which select the claims they check, produce few true verdicts, while Grok, which responds directly to questions users posed, addressed an equal share of true and false claims. 

While informative, these findings rest on queries directed to one bot on one platform over just a two month period, limiting what we can establish about how widely doubt about accurate information is shared among the broader public. Similarly, our accuracy classifications depend on Grok's answers, with human coders checking our reading of those answers and a second model providing an agreement check. The professional fact-checking comparison also describes claims that each organization selected for itself while the organizations represented in FACTors changed over time. We also did not observe whether users accepted the responses they received or changed their beliefs.

For the platforms which have either implemented or pledged to integrate crowdsourced systems like Community Notes, a practical next step would be to evaluate the public display of notes verifying posts that meet the same standards of evidence and agreement across contributors that notes flagging posts must meet. The classification filter identified in our audit of the scoring code provides a specific starting point, although allowing these notes to qualify would leave open which notes should be displayed and how their explanations should be presented. For professional fact-checkers, the equivalent question is whether claim selection can include the claims users themselves doubt. In both cases, testing whether displayed confirmations improve belief in accurate claims without encouraging confidence in content that is misleading in context would show whether these changes help users.

More broadly, given our findings on their prevalence and the availability of tools to address them, we encourage further investment in efforts to address doubted truths within wider efforts to correct public misperceptions. 

\section*{Acknowledgments}

We used Claude (Anthropic) to code the queries, Grok's replies, and the displayed notes, as described in the Methods, as well as to edit the analyses, code, and manuscript. The authors reviewed the code, verified every reported number against the analysis outputs, and take full responsibility for the paper's content.

\section*{Data and Code Availability}

The Community Notes data and scoring code are public, the latter at the commit cited. The replication repository holds the analysis code and a data release with the identifiers and coded variables for every query, post, and note we analyzed, the Perplexity and human checks, and the aggregate outputs behind each table and figure. The release omits post text, reply text, user identifiers, and handles, which can be re-fetched from the identifiers. Rerunning the full pipeline therefore requires re-fetching that text first, as the repository's documentation describes. The repository is at \url{https://github.com/morwack/true-factchecks-replication}. The code is released under the MIT License and the data release under CC BY 4.0.

\clearpage
\bibliographystyle{apacite}
\bibliography{references}
\clearpage

\appendix
\section{Display Rules}
\label{app:code}

We audited the Community Notes scoring code at commit \texttt{2704b2d} of the public repository on August 20th, 2026 \citep{cn_code}. Two rules promote notes to display, registered as \texttt{GeneralCRH} and \texttt{CRHSuperThreshold}, and both are instantiated with the parameter \texttt{onlyApplyToNotesThatSayTweetIsMisleading} set to true. The rule base class applies that filter before Helpful status is assigned:

\begin{Verbatim}[fontsize=\small]
if self._onlyApplyToNotesThatSayTweetIsMisleading:
    mask = mask & (noteStats[c.classificationKey]
      != c.noteSaysTweetIsNotMisleadingKey)
\end{Verbatim}

\noindent The three rules that reject notes set the same parameter to false, which is why they apply to both classifications, and a further rule, \texttt{NMtoCRNH}, rejects low-scoring notes verifying posts specifically. The rule instantiations sit in \texttt{note\_ratings.py} and the filter in \texttt{scoring\_rules.py}.
\clearpage

\section{Coding Instructions}
\label{app:prompts}

Each instruction below was passed as the system prompt, with the item to code in the user turn as the text of the questioned post, the query, and, where one existed, Grok's reply, each enclosed in triple quotes. The models were Claude (\texttt{claude-haiku-4-5}) for the first two instructions, Perplexity (\texttt{sonar}) for the independent check, and Claude (\texttt{claude-sonnet-5}) for the displayed notes.

\subsection{Grok's Responses}
\begin{Verbatim}[fontsize=\footnotesize]
You are coding X (Twitter) threads in which a user tagged @grok under someone's post,
  usually to ask whether it is true. You see the post being questioned, the user's
  query text, and, when Grok answered, Grok's reply. Judge stance primarily from the
  query read against the post it replies to: the query wording alone is almost always
  a bare "is this true?" and carries no signal.

Respond with a single JSON object, no prose. Keys and allowed values:

verdict: "true" | "false" | "unclear" | "refused_or_hedged"
  true: Grok affirms the questioned claim (fully or with minor caveats).
  false: Grok rejects the questioned claim (fully or mostly).
  unclear: Grok gives a mixed, partial, or non-committal substantive answer.
  refused_or_hedged: Grok declines, deflects, says it cannot verify, or answers
  something else entirely.

target: "claim_truth" | "media_authenticity" | "attribution_quote" | "other"
  claim_truth: the dispute is whether a stated fact or event is true.
  media_authenticity: the dispute is whether an image, video, or audio is real,
  AI-generated, doctored, or miscontextualized.
  attribution_quote: the dispute is whether someone really said or wrote the quoted
  words.
  other: anything else.

sourcing: "url_or_named_source" | "vague_sources" | "none"
  url_or_named_source: the reply contains a link or names a specific outlet,
  fact-checker, document, or institution.
  vague_sources: only generic references ("reports say", "sources indicate", "searches
  show").
  none: bare assertion.

Use the allowed values EXACTLY as written above. Do not shorten, abbreviate, or
invent variants: "url_or_named_source" is the literal string to return, never
"named_source". Only return the JSON object.
\end{Verbatim}

\subsection{Sincerity of the Query}
\begin{Verbatim}[fontsize=\footnotesize]
You are coding X (Twitter) threads in which a user tagged @grok as a reply to
  someone's post. You see ONLY the post being questioned and the user's query. You
  do NOT see whether or how Grok answered, and you must not speculate about it.

Judge what the user is doing. The query wording is almost always a bare "is this
  true?" and carries little signal on its own, so read it against the post it
  replies to.

Respond with a single JSON object, no prose, exactly one key:

stance: "genuine_verification" | "rhetorical_or_sarcastic" | "other"
  genuine_verification: the user genuinely asks whether the post is true, real, or
    accurate.
  rhetorical_or_sarcastic: a dunk, joke, or rhetorical challenge; the user is not
    actually uncertain (e.g. mocking the poster or Grok, baiting, a performative "is
    this true?" on obvious satire they plainly understand).
  other: not a verification request at all (opinion question, coding help, chatter),
    or unclassifiable.

Use the allowed values EXACTLY as written. Only return the JSON object.
\end{Verbatim}

\subsection{Independent Check by Perplexity}
Perplexity saw the post text, its date, and its author, and for posts with media a note that an image or video was attached.
\begin{Verbatim}[fontsize=\footnotesize]
You are a fact-checker. You will be shown a single social-media post. Determine
  whether its central factual claim is true, using web sources.

You are NOT shown, and must not speculate about, anything any AI assistant or
  fact-checker has previously said about this post. Judge the claim on the evidence
  you find.

Respond with a single JSON object, no prose:

{"core_true": "...", "verdict": "...", "confidence": "...", "rationale": "...",
  "sources": "..."}

core_true: "yes" | "no" | "cannot_determine"
  The decisive field. Would a careful reader asked "is this post's central claim
  true?" answer yes? Judge the CENTRAL claim only. Peripheral imprecision, loose
  wording, rounded figures, or missing context do NOT make it "no" so long as the
  main assertion holds. Answer "no" when the central assertion is wrong.
  "cannot_determine" only when there is no checkable central claim, or no evidence.

verdict: "true" | "false" | "mixed" | "cannot_determine"
  A secondary, stricter reading that MAY use "mixed" for a correct core with
  materially misleading framing. Never let this field change core_true.
  true: the central claim is substantially accurate.
  false: the central claim is substantially inaccurate.
  mixed: partly accurate and partly not, or accurate but materially misleading in
    context.
  cannot_determine: the post states no checkable factual claim (pure opinion, joke,
    prediction), or the claim is real but no adequate evidence is available.
    Use this freely; do not guess.

confidence: "high" | "medium" | "low"
rationale: one sentence, under 40 words.
sources: up to three source names or URLs you relied on, semicolon separated.

Only return the JSON object.
\end{Verbatim}

\subsection{Displayed notes on accurate posts}
\begin{Verbatim}[fontsize=\footnotesize]
You are shown a post from X and a Community Note that X publicly displayed beneath
  it under the label "Readers added context". The display rule means the note was
  classified by its author as indicating the post is misleading.

An independent fact-check judged the post's central claim accurate. Your task is to
  characterise what the displayed note is actually objecting to.

Respond with a single JSON object, no prose:

concedes: "yes" | "partial" | "no"
  yes: the note's own text acknowledges that the post's central claim is true or
    substantially accurate.
  partial: the note accepts part of the central claim while disputing another part
    of it.
  no: the note disputes the central claim outright, or says nothing that concedes
    it.

objection: "core_claim_false" | "peripheral_detail" | "missing_context" |
  "framing_or_implication" | "outdated" | "other"
  core_claim_false: the note asserts the post's central factual claim is wrong.
  peripheral_detail: the central claim stands and the note corrects a subsidiary
    fact.
  missing_context: the note adds context without disputing the accuracy of what was
    stated.
  framing_or_implication: the note objects to what the post implies or how it frames
    an accurate fact.
  outdated: the claim was accurate when written and has since been overtaken.
  other: none of the above.

quote: the shortest span from the note, under 20 words, that shows your reasoning.
  Empty string if none applies.

Only return the JSON object.
\end{Verbatim}
\clearpage

\section{Human Check of the Verdict Coding}
\label{app:humancheck}

Table~\ref{tab:humancheck} reports agreement between the two human coders and Claude on the 100 replies. The four-code comparisons count every reply. The true-against-false comparisons keep only the replies that both members of the pair coded true or false. Across the 100 replies, Claude coded 44 true, 34 false, 21 mixed, and 1 declined, against 48, 31, 14, and 7 for the first coder and 46, 29, 19, and 6 for the second. All three agreed on 66 replies.

\begin{table}[!htbp]
\centering
\small
\begin{tabular}{lrrrrrr}
\toprule
& \multicolumn{3}{c}{\textbf{Four codes}} & \multicolumn{3}{c}{\textbf{True against false}} \\
\cmidrule(lr){2-4}\cmidrule(lr){5-7}
& $n$ & Agreement (\%) & $\kappa$ & $n$ & Agreement (\%) & $\kappa$ \\
\midrule
Coder 1 and Coder 2 & 100 & 84.0 & 0.76 & 72 & 98.6 & 0.97 \\
Coder 1 and Claude & 100 & 74.0 & 0.60 & 68 & 97.1 & 0.94 \\
Coder 2 and Claude & 100 & 74.0 & 0.61 & 66 & 95.5 & 0.91 \\
\bottomrule
\end{tabular}
\caption{Agreement on the coding of Grok's replies, 100 random answered queries. Codes are true, false, mixed, and declined.}
\label{tab:humancheck}
\end{table}
\clearpage

\section{Agreement of Grok and an Independent Model}
\label{app:adjudication}

Table~\ref{tab:adjudication} cross-tabulates Grok's verdict with Perplexity's judgment of the post's central claim for the 600 sampled posts. The agreement rates in Section~\ref{sec:methods} are the shares of true and false verdicts that Perplexity confirmed among the posts it could resolve, 149 of 177 and 142 of 178. Among Grok's mixed verdicts Perplexity found 100 of the 167 resolved claims accurate. Perplexity reported high confidence on 415 posts, medium on 114, and low on 71, with 70 of the 71 low-confidence judgments being posts it could not resolve.

These agreement rates are higher than the 57.9\% agreement with professional fact-checkers that \citet{renault2026grok} reported for Grok-3, because the two comparisons measured different things. Theirs compared Grok's verdicts with ratings by three professional fact-checkers hired to assess 188 posts that both Grok and Perplexity had answered, using three categories (false, uncertain, true). Ours compares Grok with a second model on a stratified sample of what users asked about, and where both models err in the same direction the rates overstate agreement with the truth.

Treating Perplexity's judgments as the reference gives a true share of $p \cdot a_T + (1 - p)(1 - a_F)$, where $p$ is Grok's raw true share and $a_T$ and $a_F$ are the two agreement rates. For the top-level sample, counting each query separately, this is 50.2\%, with a 95\% interval of 46.3 to 54.3 from a parametric bootstrap that combines sampling error in the corpus with sampling error in the two validation strata. The figure moves upward because Grok's false verdicts disagreed with Perplexity more often than its true ones. One sampled record's Grok reply later failed the chronology check described in Section~\ref{sec:data}. It remains in this comparison. Table~\ref{tab:definitions} places it beside the other definitions of accurate.

\begin{table}[!htbp]
\centering
\small
\begin{tabular}{lrrrrr}
\toprule
& \multicolumn{4}{c}{\textbf{Perplexity}} & \\
\cmidrule(lr){2-5}
\textbf{Grok's verdict} & \textbf{Accurate} & \textbf{Inaccurate} & \textbf{Mixed} & \textbf{Unresolved} & \textbf{Total} \\
\midrule
True & 149 & 28 & 2 & 21 & 200 \\
False & 36 & 142 & 0 & 22 & 200 \\
Mixed & 100 & 67 & 3 & 30 & 200 \\
\midrule
Total & 285 & 237 & 5 & 73 & 600 \\
\bottomrule
\end{tabular}
\caption{Grok's verdict against Perplexity's judgment of the central claim, 600 query records on 594 posts, sampled 200 per verdict class from the earlier sincere-only subset. Unresolved covers claims with no checkable central claim or no adequate evidence. The agreement rates exclude the Unresolved and Mixed columns.}
\label{tab:adjudication}
\end{table}
\clearpage

\section{Objections Raised by Displayed notes}
\label{app:objection}

Where a displayed Community Note was published on a post Grok judged to be true, this could reflect the note and Grok simply disagreeing about underlying facts. To separate disagreement from these potential misapplications, we coded every displayed note for what it objected to using Claude (\texttt{claude-sonnet-5}). The categories include notes that assert the central claim is false, notes which object to framing or implication while the stated facts stand, notes which supply context without contesting the facts, notes which state the claim was accurate when made and has since been overtaken, notes that correct a subsidiary detail while accepting the central claim, and an Other category. A second field recorded whether the note concedes the central claim outright, in part, or not at all. Claude coded all 171 displayed notes on the 152 posts, recording the model identifier returned for the notes coded after the October 1st, 2026 release.

Since displayed notes are rare, their content matters as much as their number. The 152 posts that Grok judged true and on which a note was displayed accounted for 15.7\% of the 968 posts with a displayed note. We coded the 171 displayed notes on these posts for what they objected to (Table~\ref{tab:objection}). In 57.3\% of cases the note asserted that the post's central claim was false, which put the note and Grok in direct conflict. In 37.4\% of cases the note did not dispute the central claim but objected to its framing or implication (20.5\%), added context without contesting the facts (6.4\%), observed that an accurate statement had since been overtaken (5.8\%), or corrected a subsidiary detail (4.7\%). A further 1.8\% were coded other and 3.5\% returned no category. One such note opened by granting that ``the core story is true'' before correcting two subsidiary facts, while the label above it told every reader that readers added context. The 64 notes in these four categories appeared on 58 distinct posts, or 6.0\% of all posts with a displayed note; each of these posts therefore carried at least one note coded as not disputing its central claim.

\begin{table}[!htbp]
\centering
\small
\begin{tabular}{lrr}
\toprule
\textbf{Reasons Claimed for note} & \textbf{N} & \textbf{Share} \\
\midrule
Central claim is false & 98 & 57.3\% \\
Framing or implication, facts stand & 35 & 20.5\% \\
Missing context, facts not contested & 11 & 6.4\% \\
Claim accurate when made, since overtaken & 10 & 5.8\% \\
Subsidiary detail, central claim stands & 8 & 4.7\% \\
Other & 3 & 1.8\% \\
No category returned & 6 & 3.5\% \\
\midrule
Total & 171 & 100\% \\
\bottomrule
\end{tabular}
\caption{What the 171 displayed notes on posts Grok judged true object to, coded by Claude without a human check.}
\label{tab:objection}
\end{table}
\clearpage

\section{Ratings on notes attached to Queried Posts}
\label{app:ratings}

The three tables below use every rating in the Community Notes release of October 1st, 2026 on the 10,925 notes that contributors wrote on the queried top-level posts, and, for Table~\ref{tab:ratings_year}, on every note in the release. A rating's ``Helpful'', ``Somewhat Helpful'', or ``Not Helpful'' level is the rater's answer to the form's first question. The ``note not needed'' tag is one of the reasons a rater can give for a ``Not Helpful'' rating. The revised rating form of October 2022 offers it for notes flagging posts only, which is why its share is zero for notes verifying posts from 2023 onward. Ratings per note in 2026 are lower in every row because those notes are younger. ``Helpful'' shares are raw shares of ratings, not the bridging scores the code computes from them.

\begin{table}[!htbp]
\centering
\small
\begin{tabular}{lrr}
\toprule
& \textbf{Notes flagging} & \textbf{Notes verifying} \\
\midrule
Notes & 7,944 & 2,981 \\
Notes with any rating (\%) & 98.6 & 94.9 \\
Notes with five or more ratings (\%) & 91.6 & 82.1 \\
Median ratings per note & 56 & 50 \\
Mean ratings per note & 186.2 & 153.4 \\
``Helpful'' share of ratings (\%) & 50.0 & 60.6 \\
Not ``Helpful'' share of ratings (\%) & 47.9 & 37.3 \\
``Note not needed'', share of ``Not Helpful'' ratings (\%) & 54.3 & 0.0 \\
``Incorrect'', share of ``Not Helpful'' ratings (\%) & 29.4 & 35.0 \\
\bottomrule
\end{tabular}
\caption{Ratings received by the notes on queried top-level posts, by the note's classification.}
\label{tab:ratings}
\end{table}

\begin{table}[!htbp]
\centering
\small
\begin{tabular}{lrrrr}
\toprule
& \multicolumn{2}{c}{\textbf{Notes flagging}} & \multicolumn{2}{c}{\textbf{Notes verifying}} \\
\cmidrule(lr){2-3}\cmidrule(lr){4-5}
Post Grok judged & \textbf{False} & \textbf{True} & \textbf{False} & \textbf{True} \\
\midrule
Notes & 4,238 & 1,674 & 1,037 & 1,046 \\
Notes with five or more ratings (\%) & 91.1 & 92.1 & 86.6 & 73.5 \\
Median ratings per note & 45 & 79 & 49 & 31.5 \\
``Helpful'' share of ratings (\%) & 54.7 & 47.8 & 62.2 & 58.6 \\
Not ``Helpful'' share of ratings (\%) & 43.0 & 50.0 & 35.6 & 39.4 \\
\bottomrule
\end{tabular}
\caption{Ratings received by the notes on queried top-level posts, by the note's classification and Grok's verdict on the post. Notes on posts with a mixed or no verdict are omitted.}
\label{tab:ratings_verdict}
\end{table}

\begin{table}[!htbp]
\centering
\footnotesize
\setlength{\tabcolsep}{4pt}
\begin{tabular}{lrrrrrr}
\toprule
& \multicolumn{3}{c}{\textbf{Notes flagging}} & \multicolumn{3}{c}{\textbf{Notes verifying}} \\
\cmidrule(lr){2-4}\cmidrule(lr){5-7}
\textbf{Year} & \textbf{Notes} & \textbf{Ratings/note} & \textbf{``Helpful'' (\%)} & \textbf{Notes} & \textbf{Ratings/note} & \textbf{``Helpful'' (\%)} \\
\midrule
2021 & 14,720 & 7.7 & 45.6 & 2,427 & 10.0 & 44.7 \\
2022 & 16,314 & 40.0 & 58.0 & 2,721 & 38.5 & 43.4 \\
2023 & 275,926 & 67.5 & 66.4 & 88,637 & 48.2 & 57.0 \\
2024 & 796,001 & 76.4 & 60.5 & 250,197 & 68.8 & 59.4 \\
2025 & 658,489 & 83.5 & 63.1 & 151,298 & 78.5 & 56.8 \\
2026 & 756,993 & 49.7 & 67.1 & 119,819 & 52.4 & 54.6 \\
\midrule
\multicolumn{7}{l}{\textit{``Note not needed'', share of ``Not Helpful'' ratings (\%)}} \\
2021 & \multicolumn{3}{c}{17.4} & \multicolumn{3}{c}{4.2} \\
2022 & \multicolumn{3}{c}{49.4} & \multicolumn{3}{c}{11.2} \\
2023 to 2026 & \multicolumn{3}{c}{52.9 to 57.2} & \multicolumn{3}{c}{0.0} \\
\bottomrule
\end{tabular}
\caption{Ratings received by every note in the release of October 1st, 2026, by the note's classification and the year it was written.}
\label{tab:ratings_year}
\end{table}
\clearpage

\section{Definitions of ``Accurate''}
\label{app:definitions}

Table~\ref{tab:definitions} reports the share of posts, or of queries, that Grok judged true under each definition we considered for the top-level sample. The main definition assigns each post Grok's first true or false answer. The alternatives at the post level differed from it by at most 0.7 points, with the largest difference coming from the earlier rule under which any true answer made the post true. Counting each query separately gave 46.9\%, since posts Grok judged false drew more queries than posts it judged true.

\begin{table}[!htbp]
\centering
\small
\begin{tabular}{llrr}
\toprule
\textbf{Definition} & \textbf{Unit} & $n$ & \textbf{Judged true (\%)} \\
\midrule
First true or false answer (main definition) & posts & 19,042 & 51.4 \\
First answer of any kind & posts & 18,704 & 51.5 \\
Majority of true or false answers, ties dropped & posts & 18,898 & 51.5 \\
Posts with conflicting answers dropped & posts & 18,809 & 51.5 \\
Any true answer makes the post true (earlier rule) & posts & 19,042 & 52.1 \\
Each query counted separately & queries & 23,572 & 46.9 \\
Each query, reweighted with Perplexity as reference & queries & 23,572 & 50.2 (46.3 to 54.3) \\
\bottomrule
\end{tabular}
\caption{Share of top-level queried posts, or queries, that Grok judged true under each definition of accurate. The interval for the reweighted share is a 95\% parametric bootstrap interval.}
\label{tab:definitions}
\end{table}
\clearpage

\section{Results under Alternative Sample Definitions}
\label{app:samples}

Table~\ref{tab:samples} repeats the main results for all live queried posts including replies, for the top-level posts used throughout the paper, and for the subset of queries Claude coded as sincere. No result moves by more than two percentage points across the three definitions.

\begin{table}[!htbp]
\centering
\footnotesize
\begin{tabular}{lrrr}
\toprule
& \textbf{All live} & \textbf{Top-level} & \textbf{Top-level,} \\
& \textbf{posts} & \textbf{posts} & \textbf{sincere only} \\
\midrule
Queried posts & 34,678 & 29,294 & 25,172 \\
Queries & 42,094 & 36,675 & 31,696 \\
Queries with a recovered Grok answer & 34,459 & 30,390 & 26,449 \\
Queries with a true or false verdict & 26,406 & 23,572 & 20,868 \\
\midrule
Posts with a true or false verdict & 21,862 & 19,042 & 16,816 \\
Share of these posts judged true (\%) & 51.0 & 51.4 & 53.2 \\
\midrule
Posts with no note (\%) & 86.7 & 84.4 & 83.7 \\
Posts with any note (\%) & 13.3 & 15.6 & 16.3 \\
Posts with a displayed note (\%) & 2.8 & 3.3 & 3.4 \\
Share of notes written that verify the post (\%) & 27.4 & 27.3 & 26.2 \\
Displayed note among accurate posts (\%) & 1.4 & 1.6 & 1.6 \\
Displayed note among inaccurate posts (\%) & 6.0 & 7.0 & 7.3 \\
\bottomrule
\end{tabular}
\caption{Main results under three sample definitions. All live posts are those whose reply status the second fetch recovered, twelve fewer than the posts with text in Table~\ref{tab:corpus}. Accurate and inaccurate in the last two rows are Grok's post-level verdict under the rule in Section~\ref{sec:methods}.}
\label{tab:samples}
\end{table}
\clearpage

\section{Coverage Weighted by Reach}
\label{app:reach}

Table~\ref{tab:reach} repeats the main coverage measures with each queried post weighted by its reach, measured as the post's like count when we re-fetched the posts in August 2026. View counts were not available. Contributors wrote notes disproportionately on posts with high reach, which is why the weighted shares of written notes were about twice the unweighted shares. Displayed notes reached a much smaller share of likes. The gap between posts Grok judged true and posts it judged false persisted under weighting. Table~\ref{tab:reach_fifths} shows that the share of posts Grok judged true was similar in every fifth of the reach distribution. Two caveats apply. Likes measure engagement and serve here as a proxy for reach. They were counted in August 2026, after many of the notes had been displayed. If displayed notes reduced subsequent likes, as \citet{slaughter2025} found on average, weighting by these later likes may understate display coverage relative to engagement without the notes. The weights also describe exposure and do not establish that reach caused contributors to write.

\begin{table}[!htbp]
\centering
\small
\begin{tabularx}{\textwidth}{>{\raggedright\arraybackslash}Xrrr}
\toprule
& \textbf{Posts} & \textbf{Share of posts (\%)} & \textbf{Share of likes (\%)} \\
\midrule
All posts: any note written & 29,294 & 15.6 & 33.6 \\
All posts: note displayed & 29,294 & 3.3 & 5.3 \\
Grok judged true: any note written & 9,785 & 10.4 & 24.0 \\
Grok judged true: note flagging the post displayed & 9,785 & 1.6 & 2.1 \\
Grok judged true: note verifying the post written & 9,785 & 5.0 & 15.9 \\
Grok judged false: any note written & 9,257 & 25.7 & 49.1 \\
Grok judged false: note flagging the post displayed & 9,257 & 7.0 & 11.1 \\
Grok judged false: note verifying the post written & 9,257 & 6.6 & 21.9 \\
Posts Grok judged true, among posts with a true or false verdict & 19,042 & 51.4 & 50.7 \\
\bottomrule
\end{tabularx}
\caption{Community Notes coverage of queried top-level posts, unweighted and weighted by each post's like count.}
\label{tab:reach}
\end{table}

\begin{table}[!htbp]
\centering
\small
\begin{tabular}{rrrrrr}
\toprule
\textbf{Fifth} & \textbf{Posts} & \textbf{Median likes} & \textbf{Any note (\%)} & \textbf{Displayed (\%)} & \textbf{Judged true (\%)} \\
\midrule
1 & 5,859 & 161 & 1.2 & 0.1 & 54.9 \\
2 & 5,859 & 1,400 & 5.9 & 1.1 & 53.0 \\
3 & 5,858 & 4,394 & 14.3 & 3.8 & 50.6 \\
4 & 5,859 & 11,151 & 22.2 & 5.4 & 49.8 \\
5 & 5,859 & 36,039 & 34.4 & 6.2 & 49.1 \\
\bottomrule
\end{tabular}
\caption{Queried top-level posts divided into fifths by like count, from lowest (1) to highest (5). Judged true is the share of posts with a true or false verdict that Grok judged true.}
\label{tab:reach_fifths}
\end{table}
\clearpage

\section{Timing of notes Relative to the First Query}
\label{app:timing}

\begin{table}[!htbp]
\centering
\small
\begin{tabular}{lrr}
\toprule
& \textbf{First note written} & \textbf{First displayed note} \\
& ($n = 4{,}565$ posts) & ($n = 968$ posts) \\
\midrule
Before the first query (\%) & 45.5 & 5.7 \\
Within one hour of the query (\%) & 55.2 & 11.4 \\
Within one day (\%) & 95.9 & 82.7 \\
Within one week (\%) & 99.5 & 96.8 \\
\midrule
Hours after the first query, 10th percentile & $-$11.5 & 0.8 \\
25th percentile & $-$3.2 & 3.1 \\
Median & 0.5 & 7.3 \\
75th percentile & 4.8 & 17.4 \\
90th percentile & 14.5 & 41.7 \\
\bottomrule
\end{tabular}
\caption{Timing of the first note written and the first note displayed on a queried post, relative to the first Grok query on that post. These figures use the October 1st, 2026 release, like every other Community Notes figure. Negative hours mean the note preceded the query. Cumulative shares include notes written before the query.}
\label{tab:timing}
\end{table}
\clearpage

\section{Topic Classifier}
\label{app:topics}

The classifier scores a post's text together with its first query against two term lists per topic and counts how many of the two lists match. The topic with the highest count is assigned, ties go to the topic listed first in Table~\ref{tab:topic_terms}, and a post that matches no list is coded other or unclear. Terms match as whole words, with an asterisk marking a stem that accepts any ending.

\begin{table}[!htbp]
\centering
\footnotesize
\begin{tabularx}{\textwidth}{>{\raggedright\arraybackslash}p{3.6cm}>{\raggedright\arraybackslash}X}
\toprule
\textbf{Topic} & \textbf{Terms} \\
\midrule
War and geopolitics & war, military, army, navy, missile, nato, troops, invasion, ceasefire, sanctions, ukrain*, russi*, israel*, palestin*, gaza, iran*, china, taiwan, hamas \\
Politics and government & president, congress, senate, parliament, election, vote, democrat*, republican*, maga, trump, biden, government, governor, white house, minister, policy, immigration, border, tax, federal \\
Health and medicine & health, medical, medicine, doctor, hospital, cancer, vaccine, vaccin*, covid, virus, disease, drug, autism, diet, nutrition, patient, clinical, therapy, pharma*, fda \\
Crime, law and safety & crime, criminal, arrest(:ed), police, court, judge, lawsuit, prison, jail, fraud, scam, murder, killed, traffick*, assault, rape, shooting, terroris*, illegal \\
Economy and business & econom*, inflation, market, stock*, bitcoin, crypto*, price*, salary, jobs, unemployment, company, business, bank*, trade, tariff, budget, debt, million, billion \\
Science, climate and environment & science, scientist, climate, weather, temperature, warming, earthquake, hurricane, flood, environment, space, nasa, planet, solar, energy, carbon, species, wildlife \\
Technology and AI & ai, artificial intelligence, chatgpt, robot*, software, app*, iphone, android, tesla, internet, computer, algorithm, deepfake, cyber, hack*, social media \\
Sports & sport*, football, soccer, basketball, baseball, cricket, tennis, olympic*, fifa, nfl, nba, mlb, match, game, player, coach, goal, championship, league, team \\
Entertainment and culture & movie, film, music, song, singer, actor, actress, celebrity, tv, television, youtube, netflix, book, novel, art, festival, concert, hollywood \\
Religion and identity & religion, religious, christian*, muslim*, islam*, jew*, bible, church, god, allah, koran, quran, race, racism, gender, transgender, lgbt*, immigrant* \\
\bottomrule
\end{tabularx}
\caption{Term lists of the keyword topic classifier, in the order used to break ties.}
\label{tab:topic_terms}
\end{table}
\clearpage

\section{Display Rates by Topic and Accuracy}
\label{app:topics_full}

\begin{table}[!htbp]
\centering
\footnotesize
\begin{tabularx}{\textwidth}{>{\raggedright\arraybackslash}X*{4}{>{\centering\arraybackslash}p{1.7cm}}*{2}{>{\centering\arraybackslash}p{1.9cm}}}
\toprule
\textbf{Topic} & \textbf{Posts} & \textbf{Accurate} & \textbf{Any note} & \textbf{Displayed} & \textbf{Displayed, accurate} & \textbf{Displayed, inaccurate} \\
\midrule
War and geopolitics & 4,530 & 50.8\% & 15.5\% & 3.0\% & 1.1\% & 6.8\% \\
Politics and government & 4,378 & 52.0\% & 20.2\% & 2.7\% & 1.3\% & 5.5\% \\
Economy and business & 2,327 & 57.7\% & 11.1\% & 2.8\% & 1.1\% & 7.1\% \\
Crime, law, and safety & 1,608 & 53.2\% & 20.6\% & 3.7\% & 1.4\% & 7.8\% \\
Sports & 1,116 & 55.0\% & 11.3\% & 1.9\% & 0.9\% & 4.6\% \\
Technology and AI & 1,075 & 55.5\% & 15.2\% & 5.1\% & 3.8\% & 10.0\% \\
Health and medicine & 723 & 49.6\% & 21.7\% & 4.7\% & 3.3\% & 8.4\% \\
Religion and identity & 668 & 49.4\% & 17.4\% & 3.7\% & 1.0\% & 7.4\% \\
Entertainment and culture & 441 & 50.2\% & 15.2\% & 4.1\% & 0.7\% & 9.1\% \\
Science, climate, and environment & 322 & 58.3\% & 18.3\% & 4.3\% & 3.8\% & 8.6\% \\
Other or unclear & 12,106 & 49.3\% & 14.0\% & 3.5\% & 1.7\% & 7.2\% \\
\midrule
All & 29,294 & 51.4\% & 15.6\% & 3.3\% & 1.6\% & 7.0\% \\
\bottomrule
\end{tabularx}
\caption{Table~\ref{tab:topics} with the share of posts carrying a displayed note among those judged accurate and among those judged inaccurate. In every topic the inaccurate share exceeds the accurate share.}
\label{tab:topics_full}
\end{table}

\clearpage
\section{Monthly Composition of Community Notes}
\label{app:monthly_notes}

\begin{figure}[!htbp]
\centering
\includegraphics[width=\linewidth]{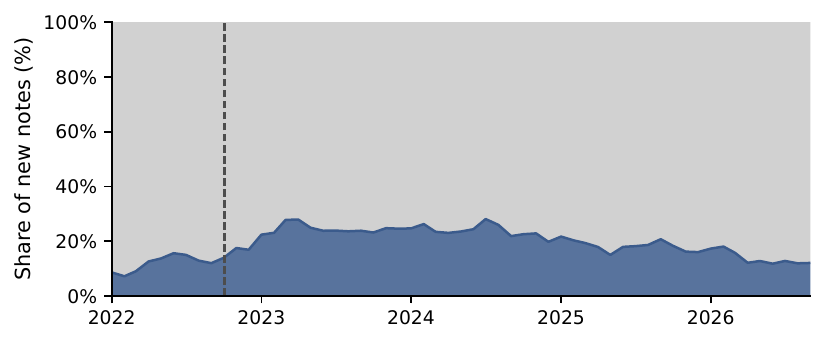}
\caption{Monthly composition of new Community Notes, January 2022 to September 2026. The blue area represents the share classified as not misleading. The dashed line marks the October 2022 revision of the rating form, after which X/Twitter resumed assigning ``Not Helpful'' status to notes verifying posts in November 2022.}
\label{fig:share_monthly}
\end{figure}

\end{document}